\documentstyle[12pt]{article}
\begin{document}
\pagestyle{empty}
\vspace*{15mm}
\begin{center}
{\large {\bf A Modification of Grover's Algorithm as a Fast Database Search}}\\
\vspace*{1cm}
{\bf D. A. Ross} \\
\vspace{0.3cm}
Department of Physics and Astronomy,\\
University of Southampton,\\
Southampton SO17 1BJ, United Kingdom\\
\vspace{0.3cm}
dar@phys.soton.ac.uk\\
\vspace{0.3cm} July 1998 \\
\vspace*{2cm}
\end{center} 

\begin{abstract} A  modification of Grover's algorithm is proposed, which can be used
directly as a fast database search. An explicit two q-bit example is displayed
in detail. We discuss the case where the database has multiple entries
 corresponding to the same target value. \end{abstract}
 
\vspace*{20mm}

\setcounter{page}{1}
\pagestyle{plain}

In the usual application of Grover's algorithm \cite{gr,gr2}, one envisages
an ``oracle'' , $T$, into which one inputs an integer $I$ and which returns
the value $T(I)=0$ for all values of $I$ except $I=I_0$ for which
 $T(I_0)=1$. A practical application of this algorithm (if it could be realised
experimentally for very large integers) could be a fast search of a database.
In other words given a function $f(I)$ which maps the integers $I$ into
the integers $F$, but which cannot be easily inverted, the algorithm
could be used to find for which value of the input integer $I$, the
function $f(I)$ was equal to $F$. In order to use the algorithm for
 this purpose it is necessary to consider the interaction of the ``database''
with the oracle, $T$, which is likely to be very involved.

In this note we propose a minor modification 
of Grover's algorithm which can be used directly for such database searches.
To begin with we will assume that the function is one-to-one and onto. We
will discuss the consequences of removing this restriction later.

 We suppose that the argument $I$ is stored in an $L$ q-bit control register
in the state $|I>$ and the value $F=f(I)$ is stored in an $L$ q-bit
target register in the state $|F>$. We consider
 a device $U_f$ which encodes the function $f(I)$ such that the action of the device
may be represented by the unitary operator $\hat{U}_f$ where
\begin{equation}
\hat{U}_f |I>\otimes|K> \ = \ |I> \otimes
 |K\oplus f(I)>.
\end{equation}
The device $U_f$ is reversible which means that the operator $\hat{U}_f$
is idempotent. In particular
\begin{equation}
\hat{U}_f |I>\otimes|0> \ = \ |I> \otimes
 |f(I)>.
\end{equation} 
\begin{equation}
\hat{U}_f |I>\otimes|f(I)> \ = \ |I> \otimes
 |0>.
\end{equation}

Other devices which we shall need are a Hadamard gate,
 $H^{(c)}$, \cite{wh} acting on the control register
whose actions on each q-bit in that register may be represented by
 the matrix
 $$\frac{1}{\sqrt{2}} \left( \begin{array}{cc} 1&1\\1&-1 \end{array} \right), $$
and the operators $S^{(c)}_I$  ( $S^{(t)}_F$) which rotate the phase of
the state by $\pi$ if the control (target) register is in the state $|I>$
($|F>$), but leaves all other states unaltered. 

Such a device could be realized in practice by the addition of an extra
q-bit for each register, called the auxiliary q-bit. Thus, for example,
for the device $S^{(t)}_{F_0}$, the target register could be 
spin-$\frac{1}{2}$ particles at $L$ different sites in a molecule or polymer 
and the auxiliary q-bit could be a further spin-$\frac{1}{2}$ particle
located at a different site. The auxiliary bit is sufficiently close to the q-bits
of the target register so that the spin-spin interactions are non-negligible,
but distinct owing to the different separations between the q-bits of
the target register and the auxiliary q-bit. On the other hand the
 spin-$\frac{1}{2}$ particles representing the control register would have to
be sufficiently far from this auxiliary q-bit for  spin-spin interactions 
between the control register q-bits and the auxiliary q-bit to be negligible.
\footnote{It will, however, be necessary for the control register q-bits
and the target register q-bits to be sufficiently close to each other so
that the spin-spin interactions between them may be employed in order to
construct the device $U_f$.} Now if the magnetic moment of the auxiliary q-bit
is $\mu$ and a  static magnetic field of magnitude $B$ is applied in
the $z$-direction, then the Hamiltonian of the auxiliary q-bit when 
it is in the state $|i>$, is
\begin{equation}
H_{aux} \ =  \  \mu \, B \left(i-\frac{1}{2}\right)
 + \sum_{l=1}^{L} \lambda_l \left(-1\right)^{(i+f_l)}, \end{equation}
where $\lambda_l$ is proportional to the interaction between the auxiliary
q-bit and the $l^{th}$ q-bit of the target register, which is in the state
 $|F>$ where $f_l$ is the $l^{th}$ bit of $F$, i.e.
 $$ F = \sum_{l=1}^{L} f_l 2^{(l-1)}.$$
The resonant frequency for spin flip of the auxiliary bit depends on
 $F$ and may be written
\begin{equation} 
\omega_{res}(F) \ =  \  \mu \, B 
 - \sum_{l=1}^{L} \lambda_l \left(-1\right)^{f_l}, \end{equation}
A $180^0$ pulse of RF magnetic field with frequency $\omega_{res}(F_0)$
in the $y$-direction will flip the
spin of the auxiliary q-bit if and only if the target register is in the 
state $|F_0>$. If we initialise the auxiliary q-bit in the state
 $$ \frac{1}{\sqrt{2}} \left( |0>-|1> \right),$$
this will introduce a minus sign if and only if the target register is
in the state $|F_0>$. The device $S^{(c)}_I$ can be constructed 
in a similar manner .

We suppose that we wish to search for the (unknown)
value $I_0$ for which $f(I_0)$ is equal to the known value $F_0$.
It is convenient to work in the two dimensional subspace of states
defined by
\begin{enumerate}
\item The state we wish to project out
 \begin{equation} |\Phi_1> \ \equiv \ |I_0> \otimes |F_0> \end{equation}
\item An orthogonal state consisting of a superposition of all 
the other states with equal coefficients
\begin{equation}
 |\Phi_2> \ \equiv \ \frac{1}{\sqrt{2^L-1}}\sum_{I\neq I_0} |I> \otimes |f(I)>
.\end{equation} \end{enumerate}

The system is initialised by acting on the state $|0> \otimes |0>$ with
the Hadamard gate on the control register followed by the device
$U_f$. This produces the state
\begin{equation}
|\Psi_1> \ =  \ \hat{U}_f \hat{H}^{(c)} |0> \otimes |0>
 \ = \ \cos \beta |\Phi_1> + \sin \beta |\Phi_2>, \end{equation}
  where $$\sin\beta \ = \ \frac{1}{\sqrt{2^L}} .$$

Now we consider the combination of operations, $S^{(t)}_{F_0}$,
followed by $U_f$, followed by $H^{(c)}$ followed by $S^{(c)}_0$
followed by $H^{(c)}$ followed by $U_f$. A little algebra shows that
in the two-dimensional subspace under consideration the operator $\hat{O}$
corresponding to this combination of operations, 
 $$ \hat{O} \ \equiv \ \hat{U}_f \hat{H}^{(c)} \hat{S}^{(c)}_0
     \hat{H}^{(c)} \hat{U}_f \hat{S}^{(t)}_{F_0},$$
may be written as the $2 \times 2$ matrix, \cite{bbht,cz}
 \begin{equation}
\hat{O} \ = \ - \left( \begin{array}{cc} \cos 2\beta & \sin 2 \beta \\
-\sin 2 \beta & \cos 2 \beta \end{array} \right). \label{eqo} \end{equation}
This is most readily seen by observing that the operator $\hat{S}^{(c)}_0$
(acting on the control register) may be written
\begin{equation}
\hat{S}^{(c)}_0 \ = \ I - 2 |0><0|, \end{equation}
($I$ is the identity), whereas
the operator $\hat{S}^{(t)}_{F_0}$ (acting on the target register)
may be written
\begin{equation}
\hat{S}^{(t)}_{F_0} \ = \ I - 2 |F_0><F_0|. \end{equation}
Together with the matrix elements
\begin{equation}
 <\Phi_1|\hat{U}_f \hat{H}^{(c)} |0> \otimes|0>= \sin\beta \end{equation}
\begin{equation}
 <\Phi_2|\hat{U}_f \hat{H}^{(c)} |0> \otimes|0>= \cos\beta, \end{equation}
the result (\ref{eqo}) follows.

Thus we see that the application of the operator $\hat{O}$ $N$ times
where $N$ is the nearest integer to the quantity $\nu$ where
 \begin{equation}
 \nu \ = \ 
\frac{\pi}{4\sin^{-1}\left(\frac{1}{\sqrt{2^L}}\right)}-\frac{1}{2}
 \label{eqnu} \end{equation}
 on the initial state $|\Psi_1>$
will project the system into a state which is almost pure $|\Phi_1>$.
A measurement of the control register will now return the value $I_0$.
At the same time one should measure the state of the target register
as a check. Errors can occur owing to the fact that there is still
some probability  (of order $1/2^L$)
for the system to be in the unwanted state $|\Phi_2>$.
These errors are, of course, amplified by decoherence effects. However, the
error rate should still be small compared with unity so that any error
is very unlikely to survive a repetition of the experiment.  

 As a pedagogical exercize we now consider a simple example in which the control and target registers
each consist of two q-bits. Note that if we set $L=2$ in eq.(\ref{eqnu})
we find exactly $\nu=1$, which means that the system should be
 in the required state after a single pass through the combination of devices.
The standard Grover's algorithm for two q-bits has been achieved
using NMR \cite{cc,jj} and this is currently being extended to three q-bits,
 so perhaps the example considered here presents
the next experimental challenge.
We will take the example of the function
 \begin{equation} f(I)=(3-I), \ \ (I=1, \cdots 3), \end{equation}
and pretending that we are unable to invert this function we search
for the value of $I$ for which $f(I)=2$.

For two q-bit registers the devices required are represented by operators which
can be written in terms of $2 \times 2$ matrices. For the operators acting
on the control register we have
\begin{equation}
 \hat{H}^{(c)} \ = \ \frac{1}{2}\left( \begin{array}{cccc}
  1&1&1&1\\1&-1&1&-1\\1&1&-1&-1\\1&-1&-1&1 \end{array} \right) \end{equation}
\begin{equation}
 \hat{S}^{(c)}_0 \ = \ \left( \begin{array}{cccc}
  -1&&&\\&1&&\\&&1&\\&&&1 \end{array} \right). \end{equation}
The operator acting on the target register is given by
\begin{equation}
 \hat{S}^{(t)}_2 \ = \ \left( \begin{array}{cccc}
  1&&&\\&1&&\\&&-1&\\&&&1 \end{array} \right), \end{equation}
and the operator $\hat{U}_f$ acts on both registers. We shall not write this
out  explicitly. The only action that we need is
\begin{equation}
\ \hat{U}_f |I>\otimes |0>
  \ = \ |I>\otimes|3-I> . \end{equation}

Now we follow through the algorithm step by step.   
\begin{enumerate}
\item Pass the system in the state  $|0>\otimes|0>$  through a
 Hademard gate
on the control register to obtain
\begin{equation}
 |\Psi_0> \ = \ \hat{H}^{(c)}|0> \otimes |0> \ = \ \frac{1}{2}
\left( |0>\otimes|0>+|1>\otimes|0> + |2>\otimes|0>
  +|3>\otimes|0>\right). \end{equation}
\item Pass the system in the state $|\Psi_0>$ through the device
 $U_f$ to obtain
 \begin{equation} |\Psi_1> \ = \ \hat{U}_f|\Psi_0>  \ = \ \frac{1}{2}
\left( |0>\otimes|3>+|1>\otimes|2> + |2>\otimes|1>
 +|3>\otimes|0>\right). \end{equation}
\item
Pass the system in the state $|\Psi_1>$ through the device $S^{(t)}_2$
to obtain
 \begin{equation} |\Psi_2> \ = \ \hat{S}^{(t)}_2|\Psi_1>  \ = \ \frac{1}{2}
\left( |0>\otimes|3>-|1>\otimes|2> + |2>\otimes|1>
  +|3>\otimes|0>\right). \end{equation}
 \item Pass the system in the state $|\Psi_2>$ through the device $U_f$
to obtain 
 \begin{equation} |\Psi_3> \ = \ \hat{U}_f|\Psi_2>  \ = \ \frac{1}{2}
\left( |0>\otimes|0>-|1>\otimes|0> + |2>\otimes|0>
 +|3>\otimes|0>\right). \end{equation}
\item Pass the system in the state $|\Psi_3>$ through the Hademard gate
acting on the control register to obtain
\begin{equation}
|\Psi_4> \ =  \ \hat{H}^{(c)}|\Psi_3> \ = \  \frac{1}{2} 
\left( |0>\otimes|0>+|1>\otimes|0> - |2>\otimes|0>
 +|3>\otimes|0>\right). \end{equation}
\item Pass the system in the state $|\Psi_4>$ though the device $S^{(c)}_0$
to obtain
\begin{equation}
|\Psi_5>  \ = \ \hat{S}^{(c)}_0|\Psi_4>  \ =  \ \frac{1}{2} 
\left(- |0>\otimes|0>+|1>\otimes|0> - |2>\otimes|0>
 +|3>\otimes|0>\right). \end{equation}
\item Pass the system in the state $|\Psi_5>$ through the Hademard gate
acting on the control register to obtain
\begin{equation}
|\Psi_6> \ = \  \hat{H}^{(c)}|\Psi_5> \  = \  -|1>\otimes|0>. 
\label{pen} \end{equation}
\item Finally we pass the system in the state $|\Psi_6>$ through the device
$U_f$ to obtain the required result
\begin{equation} |\Psi_7>  \ = \  \hat{U}_f|\Psi_6> \ = \ -|1>\otimes|2> \  (= -|\Phi_1>). \end{equation}
\end{enumerate}
Here we see role played by the modification to Grover's original method,
 namely the interposition of the device $U_f$ before the
 application of the Hademard
gate for the first time. This clears the target register and without it
the penultimate step (7) would not have led to the pure state
 $|\Psi_6>$. A further  application of $U_f$  is necessary before
the process can be iterated.

We end by discussing the situation where the function $f(I)$ is not one-to-one
or onto. Firstly suppose that the target register is larger than the control
register and we accidentally search for a value of $F_0$ to which there is no
corresponding $I_0$ such that $f(I_0)=F_0$.
 In this case the operator $\hat{S}^{(t)}_{F_0}$ becomes effectively the identity
operator and the effect of the product $\hat{O}$ of operators is just to flip
the overall sign. In this case the resulting measurement of the control
register after $N$ of these operations will simply return any one of the values of
$I$ with equal probability and there will, of course, be an error when the
target register is measured, which will persist through subsequent repetitions.

On the other hand if the function is not one-to-one then one may adapt the
 generalization proposed in refs.\cite{bbht,bbgl1,bbgl2}.
 If there are $g$ values of $I$, namely $I_0^r, \ r=1\cdots g$,
for which $f(I_0^r)=F_0$, then the number of applications of the operator
 $\hat{O}$ required to produce the required result becomes
 the nearest integer to $\nu(g)$ where  
 \begin{equation}
 \nu(g) \ = \ 
  \frac{\pi}{4\sin^{-1} \left( \sqrt{\frac{g}{2^L}} \right)}-\frac{1}{2}.
 \label{eqnug} \end{equation}
This projects out the state
\begin{equation} |\Phi_1> \ =
 \frac{1}{\sqrt{g}}\sum_{r=1}^g  |I_0^r>\otimes|F_0>,\end{equation} 
so that  a measurement of the control register will yield one of the
values of $I_0^r$ with equal probability. If the degeneragy, $g$
is unknown then the method proposed in ref.\cite{bbht} can also
be used here.
\bigskip

\noindent{\bf Acknowledgements:} The author is grateful to Tony Hey,
Richard Hughes  and Jonathan Jones for useful discussions.

\end{document}